\documentclass{article}

\usepackage{arxiv}

\usepackage[utf8]{inputenc} 
\usepackage[T1]{fontenc}    
\usepackage{hyperref}       
\usepackage{url}            
\usepackage{booktabs}       
\usepackage{amsfonts}       
\usepackage{nicefrac}       
\usepackage{microtype}      
\usepackage{graphicx}
\usepackage{doi}

\usepackage{authblk}
\usepackage{amsmath, bm}
\usepackage{upgreek}
\usepackage{longtable}
\DeclareFontFamily{OT1}{pzc}{}
\DeclareFontShape{OT1}{pzc}{m}{it}{<-> s * [1.500] pzcmi7t}{}
\DeclareMathAlphabet{\mathpzc}{OT1}{pzc}{m}{it}

\usepackage{xfrac}
\usepackage{commath}
\usepackage[title]{appendix}

\graphicspath{ {./figure/} }

\title{Phase-Field DeepONet: Physics-informed deep operator neural network for fast simulations of pattern formation governed by gradient flows of free-energy functionals}

\author{Wei Li\textsuperscript{ a, b} 
        \quad  Martin Z. Bazant\textsuperscript{ b, c}
        \quad Juner Zhu\textsuperscript{ a,} \thanks{Corresponding author.
Emails: we.li@northeastern.edu (W.L.), bazant@mit.edu (M.Z.B), j.zhu@northeastern.edu (J.Z.)}
        \\
        \textsuperscript{a}Department of Mechanical and Industrial Engineering, Northeastern University \\
        \textsuperscript{b}Department of Chemical Engineering, Massachusetts Institute of Technology\\
        \textsuperscript{c}Department of Mathematics, Massachusetts Institute of Technology}

\begin{document}

\maketitle

\begin{abstract}
    Recent advances in scientific machine learning have shed light on the modeling of pattern-forming systems. However, simulations of real patterns still incur significant computational costs, which could be alleviated by leveraging large image datasets. Physics-informed machine learning and operator learning are two new emerging and promising concepts for this application. Here, we propose ``Phase-Field DeepONet", a physics-informed operator neural network framework that predicts the dynamic responses of systems governed by gradient flows of free-energy functionals. Examples used to validate the feasibility and accuracy of the method include the Allen-Cahn and Cahn-Hilliard equations, as special cases of reactive phase-field models for nonequilibrium thermodynamics of chemical mixtures. This is achieved by incorporating the minimizing movement scheme into the framework, which optimizes and controls how the total free energy of a system evolves, instead of solving the governing equations directly. The trained operator neural networks can work as explicit time-steppers that take the current state as the input and output the next state. This could potentially facilitate fast real-time predictions of pattern-forming dynamical systems, such as phase-separating Li-ion batteries, emulsions, colloidal displays, or biological patterns. 

\end{abstract}


\keywords{Physics-informed machine learning \and deep operator neural network \and phase-field method \and minimizing movement scheme \and Allen-Cahn and Cahn-Hilliard equations}

\section{Introduction}

Machine learning (ML) has achieved enormous success in many disciplines, especially computer vision \cite{voulodimos2018} and natural language processing \cite{young2018}. 
However, when it comes to scientific problems,  ML algorithms are often thought of as black boxes that can hardly be interpreted and lack rigorous justifications by physical laws.
Recently, a concept, scientific machine learning (SciML), emerged and has quickly attracted wide attention \cite{baker2019}. Its aim is to make the traditional ML domain-aware, interpretable, and robust. 
In some sense, SciML is now revolutionizing the area of computational science and has been applied in various scientific disciplines, such as ML-enhanced multiphysics and multiscale modeling \cite{cai2021, lin2021,yin2022}, ML-assisted fast online prediction and guided data acquisition \cite{jiang2021}, and optimal decisions for complex systems \cite{jiang2022fast,jiang2021}. 
Among various SciML architectures, physics-informed machine learning (PIML) and operator learning are the two representative examples.
PIML introduces known physics into an ML algorithm and as a result, requires a smaller dataset or sometimes even works without experimental data \cite{karniadakis2021}. PIML is a general concept and can be implemented in various strategies \cite{finegan2021}. 
Considering the three ingredients of a general ML algorithm, physical laws can be accordingly incorporated into 1) the training data, e.g., data generated from first-principle-based simulations, 2) the ML models, e.g., neural networks designed to reflect certain physical principles (symmetry, positive definiteness, hierarchical relations, etc.), and 3) the training strategies, e.g., loss functions formulated to include physical laws. 
Among these different approaches, one of the most extensively studied is the physics-informed neural network (PINN) \cite{raissi2019, raissi2018, lu2019}, where the known physics, namely, the governing equations, initial conditions (ICs), and boundary conditions (BCs), are incorporated into the loss function in the form of residuals. 
It has also been demonstrated that PINNs are able to deal with both forward (solving equations) and inverse (identifying parameters) problems for fluid dynamics governed by Navier-Stokes equations \cite{raissi2019}. 
Since proposed, PINNs have been widely adopted in various applications. Interested readers can refer to \cite{karniadakis2021} for a comprehensive review.
In addition to these successful applications, PINNs have also been extended to accommodate irregular \cite{gao2021} and multiple domains \cite{wang2022cenn, jagtap2020conservative}, enforce hard constraints with modified NNs \cite{lu2021b}, incorporate adaptive activation \cite{jagtap2020}, introduce gradient-enhanced \cite{yu2022} or energy-based terms into loss function \cite{li2021physics_guided, samaniego2020}, etc.

Despite the great success, a prominent challenge of PINNs is to efficiently determine or optimize the hyperparameters in the loss function. Most existing studies applied a trial-and-error scheme which makes the training time-consuming.
To tackle this challenge, Psaros et al. \cite{psaros2022} recently proposed a meta-learning framework to optimize the loss function offline.  
Meanwhile, it is also possible to avoid this issue by reducing the total number of loss terms. One common way is to enforce the hard constraints \cite{lu2021b}, which entails modifying the neural networks such that the outputs always satisfy certain BCs and ICs. 
Another approach is to make use of the energy or variational principles that intrinsically include the PDEs and BCs. 
For example, solving the Laplacian equation $\nabla^2u = 0$ with a Neumann-type BC ${\partial_{\mathbf{n}} u}= 0$ using PINN needs at least two loss terms. 
Alternatively, this problem is equivalent to finding the minimum of the  functional $\mathcal{J} = \int_{\Omega} 0.5 (\nabla u)^2 \dif V$ (i.e. $\delta \mathcal{J} = 0$), which can be treated as the only loss term. 
In this way, the number of loss terms can be reduced.
Mathematically, this example is just a special case of the more general Euler-Lagrangian equation (see \ref{sec:appendix_EL}), and note that the order of derivatives in the functional is lower than that in the PDEs, which further improves the training efficiency.


A few existing studies have explored the strategy of using energy as the loss function. E et al. \cite{weinan2018deep} proposed a DeepRiz neural network to solve the Poisson's equation ($-\nabla^2u(x) = f(x) $) with homogeneous essential BC ($u(x) = 0, x\in \partial \Omega$) by minimizing the functional $\mathcal{J} = \int_{\Omega} \left[ 0.5 \cdot (\nabla u)^2 - f(x) \cdot u(x) \right] \dif x$, which is also a special case of the Euler-Lagrangian equation.
Later, Wang et al. extended this framework to consider inhomogeneous essential BCs at complex boundary geometries and multiple domains \cite{wang2022cenn}. 
Another case explored is the principle of minimal potential energy in solid mechanics, which states that the deformation of a solid domain will follow the path that minimizes the total potential energy under external loads \cite{samaniego2020, li2021physics_guided}. 
For quasi-static elastic responses, the total potential energy $ \mathcal{T} $ consists of the elastic strain energy $ U = \int_{\Omega} 0.5 \mathbf{\sigma}: \mathbf{\varepsilon} \dif V$ and the work potential $ W = - \int_{\partial \Omega} f \cdot u  \dif S $.
Minimizing $ \mathcal{T} = U + W$ ($\delta\mathcal{T} = 0$) is equivalent to solving the corresponding Euler-Lagrangian equation (force equilibrium) $ \nabla \cdot \mathbf{\sigma} = 0$ with BC $\mathbf{\sigma} \cdot \mathbf{n} = 0 $.
In the authors' previous work, we implemented this principle by introducing the potential energy into the loss function and predicted the deformation of elastic plates \cite{li2021physics_guided}.
We also compared this energy-based framework with the vanilla residual-based PINN and found that the energy-based one is more efficient in terms of training time due to the lower order derivatives and fewer hyperparameters in the loss function, though the accuracy of both is comparable. 
It should be mentioned that the residual-based PINN is a more universal framework, while the energy-based one is only limited to systems governed by energy or variational principles. 
To the authors' best knowledge, the above existing studies only explored simple linear systems under quasi-static conditions. 
In this study, we aim to look into the dynamics of highly nonlinear and coupled energy storage systems and make use of the variational principles to construct a PIML framework. 

Operator learning is another concept that has emerged as a promising SciML technique; it learns the mapping from one function to another function, such as the sequence-to-sequence and image-to-image mappings. 
As a comparison, many widely-used network architectures, such as fully-connected neural networks (FNNs) and convolutional neural networks (CNNs),  are finite-dimensional operators that map one discretized signal or image to another.
Recently, some novel architectures have been proposed to learn the infinite-dimensional mappings, such as Deep Operator Networks (DeepONets) \cite{lu2021} and Fourier Operator Networks (FNOs) \cite{kovachki2021}. 
A comprehensive review of the operator learning with neural networks could be found in \cite{kovachki2021}.
In this study, we will focus on the DeepONet approach developed by Lu et al. \cite{lu2021}.
So far, DeepONets have been proven to have better approximation and generalization abilities than FNNs and, therefore, has been used by many applications.
One of the advantages of DeepONets is their ability to take the BCs or ICs as inputs, making it theoretically possible to train one network for all scenarios. This means that once the network is trained, it can be used to solve new problems with different boundary and/or initial conditions without additional training. This can be particularly useful in applications where the boundary and/or initial conditions may vary or come with significant uncertainty.

These new advances in SciML have shed light on the modeling of energy-storage systems (ESSs), especially Li-ion batteries. 
Due to the high-dimensional (e.g., multiple materials, scales, and physical fields) nature of this type of systems, it is cumbersome to develop a complete physics-based model or fully interpret a big dataset.
Developing a unified physics-informed machine learning computational framework that can combine the partially-known physics and a small-size dataset is very appealing and necessary. The fundamental challenge here is the trade-off between the abundance of data and the adequacy of physical laws. At the electrode or cell level, experimental data is relatively easy to be obtained but physics is mostly hidden behind the data. Therefore, purely data-driven machine learning algorithms can be applied to predict the performance \cite{jiang2021} and lifetime \cite{severson2019} of batteries.
When data is expensive or limited, for example, battery degradation data diromg thousands of cycles taking years to complete, some studies proposed frameworks based on PINNs to estimate the states of battery cells \cite{li2021physics, tian2022battery}, identify battery parameters \cite{wu2022physics},  predict the lifetime \cite{aykol2021perspective}, and recognize degradation patterns  \cite{nascimento2021hybrid, bills2020universal, chen2021machine} at the electrode and cell levels.
On the other hand, experimental data is expensive and difficult to collect at the active particle level although many fundamental electro-chemo-mechanical physical theories have been developed at the micro-scales~\cite{Lim2016,Li2014b}.
Physics-based models is often used to describe the single-particle pattern formation and extrapolate to porous electrodes~\cite{bazant2013theory,Ferguson2012,Smith2017}, but it is often very time-consuming to solve the models. 
As explained previously, PIML has a potential advantage to produce efficient surrogates or reduced-order models due to the fast inference speed of machine learning algorithms after training. For example, several studies used PINNs to solve the two  equations for the phase-field method, namely Allen-Cahn and Cahn-Hilliard \cite{wight2020solving, mattey2021physics, mattey2022novel}, which will be elaborated on in Section \ref{sec:theory}.

Another reason for applying PIML in ESSs is that the determination of constitutive relations and material constants is challenging. Many advanced algorithms have been developed for the interpretation of large datasets of full-field image data, in the context of phase-field models for electrochemical nonequilibrium thermodynamics~\cite{bazant2013theory}. For example, Zhao et al. used PDE-constrained optimization to learn the physics of driven phase separation in lithium iron phosphate nanoparticles from operando images of scanning tunneling x-ray microscopy~\cite{zhao2022learning}.
Deng et al. used similar methods to learn the constitutive law of the eigen strain change with respect to lithium intercalation from micro X-ray tomography and diffraction images of active particles \cite{deng2022correlative}. This optimization process is often time-consuming and PIML has the potential to achieve faster identification. 

Nowadays, the greater scientific community has recognized the value of integrating physics and data into one unified framework as a high-level vision. The energy storage community is one of the pioneering areas. For example, the U.S. Department of Energy (DOE) is the first federal agency to propose the concept of SciML \cite{baker2019}. The current remaining challenge is to find realistic ways to implement them. It is always crucial to first understand the physics in order to construct a proper machine learning architecture for the studied system. The above-mentioned Allen-Cahn and Cahn-Hilliard equations are essential in chemical system modeling. They are able to describe the dynamics of non-conserved and conserved order parameters, respectively, in terms of variationally defined chemical potentials. Both equations can be derived through variational methods that have been well established to obtain the governing equations of complex coupled nonlinear systems \cite{cahn1994}. More specifically, Allen-Cahn and Cahn-Hilliard equations are two special cases of gradient flows that entail finding and constructing an appropriate free energy and an inner product to incorporate the kinetics into a variational framework \cite{cahn1994,taylor1994,santambrogio2016}. Gradient flows can be applied to a large variety of physics including diffusion, phase separation, microstructure evolution, etc. Therefore, constructing a machine learning framework for gradient flows can be beneficial to a wide range of applications.
As we mentioned earlier, ML can be adopted naturally to solve variational problems, where we can approximate the solutions with ML models by minimizing the free energy functional as a loss function. 

In this study, we propose the idea of ``Phase-Field DeepONet" as a general neural network framework for dynamical systems governed by gradient flows of free energy functionals, taking advantage of the energy-based loss function, deep operator network, and physics-informed learning. The paper is organized as follows: Section \ref{sec:theory} presents the theory of phase-field method and gradient flows; Section \ref{sec:ML_framework} describes the framework of Phase-Field DeepONet that incorporates the minimizing movement scheme into a physics-informed deep operator neural network; 
In Section \ref{sec:numerical_examples}, we investigate three different dynamical systems including the linear relaxation kinetics, Allen-Cahn, and Cahn-Hilliard dynamics to validate the proposed framework. 

\section{phase-field method and gradient flows}\label{sec:theory}

\subsection{Phase-field method}

Phase-field methods are widely used in materials science because of their capability to track microstructure evolution, grain growth and coarsening, crack propagation, etc. Unlike other sharp interface models, phase-field models treat interface in a diffusive way with phase-field variables, which can then describe the domain and all interfaces continuously as a whole. There are two types of phase-field variables, namely conserved and non-conserved fields. The evolution (or dynamics) of both are governed by the total free energy $\mathcal{F}$ of a system and its variational derivatives with respect to the field variables, which can be viewed as diffusional chemical potentials. 

For a single field variable $\phi$, the standard free energy functional for an inhomogeneous system, proposed by Van der Waals~\cite{rowlinson1979translation} and Cahn and Hilliard~\cite{cahn1958}, is defined as, 
    
\begin{equation}
    \label{eq: free energy}
    \mathcal{F} = \int_\Omega {[ f(\phi) + \frac{1}{2} \kappa_{\phi} (\nabla \phi)^2 ]} \dif x,
\end{equation}

\noindent
where $f(\phi)$ is the homogeneous free-energy density and the second term on the right-hand side represents the gradient energy at phase interfaces with the gradient coefficient $\kappa_\phi$. 
Generally, the field variables evolve in the direction where the free energy continuously decreases. 
For a conserved field variable, the dynamics can be expressed as a conservation law for gradient-driven fluxes. The Cahn-Hilliard equation can be then obtained,

\begin{equation}
    \label{eq: cahn-hilliard equation}
    \dpd{\phi}{t} = \nabla \cdot M \nabla \frac{\delta\mathcal{F}}{\delta\phi}, 
\end{equation}

\noindent
where $M$ is a transport coefficient (the product of the mobility and the concentration field variable~\cite{bazant2013theory}) and the functional derivative (diffusional chemical potential) is given by,

\begin{equation}
    \label{eq:energy_functional_derivative}
    \frac{\delta\mathcal{F}}{\delta\phi} = \dpd{f}{\phi} - \kappa_{\phi} \nabla \cdot \nabla \phi.
\end{equation}

For a non-conserved field variable, we have the Allen-Cahn equation,

\begin{equation}
    \label{eq: allen-cahn equation}
    \dpd{\phi}{t} = - M \frac{\delta \mathcal{F}}{\delta \phi},
\end{equation}

 \noindent
 where the functional derivative is also given by Eq.~\ref{eq:energy_functional_derivative}. The Allen-Cahn equation can be viewed as a linearized model of a reaction producing the field variable, proportional to the affinity, or  difference in diffusional chemical potential with respect to an external reservoir~\cite{bazant2013,Bazant2017}.
 
\subsection{Mathematics of gradient flows}

One common way to establish the theories or governing equations of a system is to start with constitutive relations based on experimental data. 
These theories need then to be checked for consistency with thermodynamic laws. 
The variational methods, on the other hand, start with thermodynamics so that the derived theories are always consistent with the thermodynamic laws. 
In addition, the variational methods can handle extreme anisotropy, non-differentiability, and nonlinearity more easily. 
In this section, we will review a general mathematical framework to derive phase-field models as gradient flows of free energy functionals. 

The second law of thermodynamics states that the total free energy of a system always decreases. 
Therefore, the equations governing the evolution of field variables in a system should be constructed in an appropriate way to guarantee a monotonic decrease in total free energy. 
One approach is the gradient flows which can be described by,

\begin{equation}
    \label{eq:gradient_flow}
    \dod{}{t} u(x,t) = - \nabla \mathcal{F}(u), 
\end{equation}

\noindent where $u$ is a field variable that evolves with time (depending on space and time); 
$\mathcal{F}$ is a smooth and convex energy functional of $u$. 
$\nabla F(u)$ indicates the functional gradient of $\mathcal{F}$ with respect to $u$. 
Physically, this equation states that the field variable $u$ evolves in the direction where the free energy decreases fastest.
The functional gradient is the driving force of the dynamic process. Mathematically, analogous to the directional derivative of a multi-variable function, the functional gradient $\nabla \mathcal{F}$ and functional derivative $\delta \mathcal{F}/ \delta u$ are related by the inner product $<.,.>$, 

\begin{equation}
    \label{eq:functional_gradient}
     <\nabla \mathcal{F}, v> = \dod{\mathcal{F} (u+v \cdot t)}{t} |_{t=0} = \int_\Omega {\frac{\delta \mathcal{F}}{\delta u} v} \dif x,
\end{equation}

\noindent where $v$ is an arbitrary function and can be viewed as a flow field; $t$ is a short time and $v \cdot t$ is also called the variation of $u$.

Therefore, the functional gradient depends on not only the free energy functional (its functional derivative) but also the construction of the inner product (or the measure of distance).

For a functional energy defined as $ \mathcal{F} = \int_\Omega {F(x, u(x), \nabla u(x)} \dif x $, the functional derivative can be determined by (see \ref{sec:appendix_func_grad})

\begin{equation}
    \label{eq:functional_derivative}
    \frac{\delta \mathcal{F}}{\delta u} = \dpd{F}{u}  - \nabla \cdot \dpd{F}{\nabla u}.
\end{equation}

In order to get the functional gradient, we still need to construct an inner product. In this study, we introduce two different inner products. The first one is the weighted $L^2$ inner product,

\begin{equation}
    \label{eq: L2 inner product}
    <f, g>_{L^2,M} = \int_\Omega {\frac{f(x) \cdot g(x)}{M}} \dif x,
\end{equation}

\noindent where $M$ is the weight; $f(x)$ and $g(x)$ are two arbitrary functions. The corresponding $L^2$ norm is

\begin{equation}
    \label{eq:L2_norm}
     |f|_{L^2,M}=\sqrt{\int_\Omega \frac{f^2(x)}{M} \dif x}.
\end{equation}

The second one is the $H^{-1}$ inner product with a weight, defined as
\begin{equation}
	\label{eq: H-1 inner product}
    <f, g>_{H^{-1},M} = \int_\Omega {\nabla \phi_f \cdot M \nabla \phi_g} \dif x,
\end{equation}

\noindent where $\phi_f$ is the solution of the Poisson's equation with Neumann boundary condition, 

\begin{equation}
	\label{eq: Poisson's_Neumann}
    \begin{cases}
        \nabla^2 \phi_f = f(x), & \text{in } \Omega, \\
        \partial_{\boldsymbol{n}} \phi_f = 0, & \text{on } \partial \Omega.
    \end{cases}
\end{equation}

Note that $\phi_f$ has an unique solution when $\int_\Omega f \dif x = \int_\Omega \phi_f \dif x = 0 $. The corresponding $H^{-1}$ norm can be obtained with

\begin{equation}
    \label{eq:H-1_Norm}
    |f|_{H^{-1},M} = \sqrt{\int_\Omega \nabla \phi_f \cdot M \nabla \phi_f \dif x}. 
\end{equation}
 
We can then obtain the functional gradient. For $L^2$ inner product, with Eq.~\ref{eq:functional_gradient} and Eq.~\ref{eq:L2_norm} we have

\begin{equation}
    \label{eq: L2 gradient}
    <\nabla \mathcal{F}, v>_{L^2,M} = \int_\Omega \frac{\nabla \mathcal{F} v}{M}  \dif x = \int_\Omega \frac{\delta \mathcal{F}}{\delta u} v \dif x.
\end{equation}

Since $v$ is an arbitrary function, to ensure the equality is always satisfied, we have,

\begin{equation}
    \label{eq:L2_Allen-Cahn}
    \nabla \mathcal{F} = M \frac{\delta \mathcal{F}}{\delta u}.
\end{equation}

Similarly, with  $H^{-1}$ inner product (Eq.~\ref{eq:functional_gradient} and Eq.~\ref{eq:H-1_Norm}) we have,

\begin{equation}
    \label{eq:H-1_Cahn-Hilliard}
    \nabla \mathcal{F} = -\nabla \cdot M \nabla \frac{\delta \mathcal{F}}{\delta u}.
\end{equation}

Substituting the above two functional gradients (Eq.~\ref{eq:L2_Allen-Cahn} and Eq.~\ref{eq:H-1_Cahn-Hilliard}) into Eq.~\ref{eq:gradient_flow}, we get the Allen-Cahn and Cahn-Hilliard equations, respectively. 

\subsection{The minimizing movement scheme}

We have demonstrated a mathematical way to derive the phase-field equations (i.e., Allen-Cahn and Cahn-Hilliard equations) with the gradient flows theory.
In order to predict the dynamic response of a system governed by gradient flows, we can directly solve these equations with the finite element (FE) or finite difference (FD) method. 
Alternatively, we can make use of an important feature of gradient flows. 
Given a fixed small time step $\tau > 0$ and a smooth and convex functional $\mathcal{F}(u)$, we can find a time sequence of $u$, $[u_1^\tau, u_2^\tau, \dots, u_n^\tau]$ from the initial condition $u_0^\tau$, through the following iterated scheme, which is called minimizing movement scheme,

\begin{equation}
    \label{eq: Min. Move. Scheme}
    u_{k+1}^\tau \in \text{argmin}_u \left[ \mathcal{F}(u) + \frac{d^2(u, u_k^\tau)}{2 \tau} \right],
\end{equation}

\noindent where $u_k^\tau$ and $u_{k+1}^\tau$ represent the field distribution at time step $k$ and $k+1$; $d(.,.)$ indicates the distance between two functions. 
The time sequence obtained from this iterative minimization scheme is actually the solution of the corresponding PDEs.
For the above minimization problem, we know $u_{k+1}^\tau$ is the solution of

\begin{equation}
    \label{eq: Min. Solution}
    \dpd{}{u} \left[ \mathcal{F}(u) + \frac{d(u, u_k^\tau)}{2 \tau} \right] = \nabla \mathcal{F}(u) + \frac{d(u, u_k^\tau)}{\tau} = 0,
\end{equation}

\noindent which gives

\begin{equation}
    \label{eq: Min. Solution 2}
     \frac{d(u, u_k^\tau)}{\tau} = - \nabla \mathcal{F}(u).
\end{equation}

This equation is the discrete-time implicit Euler scheme for Eq.~\ref{eq:gradient_flow}. Therefore, we can get the solution by the minimizing movement scheme instead of solving the PDEs directly. 

\section{Phase-Field DeepONet: physics-informed deep operator neural network for gradient flows }\label{sec:ML_framework}

In this section, we propose Phase-Field DeepONet, a physics-informed deep operator neural network framework incorporating the aforementioned minimizing movement scheme to solve the gradient flows of free energy functionals. 
There are two important ingredients in this framework: 1) operator learning, which aims to learn the mapping from one function to another function. 
In this study, we make use of this concept to learn the mapping of the field variable distribution from the current time step to the next time step; 
2) physics-informed machine learning, which incorporates known physics into a machine learning framework. 
We directly utilize Eq.~\ref{eq: Min. Move. Scheme} as the loss function to implement the physics of gradient flows. 
These two aspects will be further explained in the following.

\subsection{Deep Operator Neural Network (DeepONet)}

\begin{figure}
    \centering
    \includegraphics[width=\textwidth]{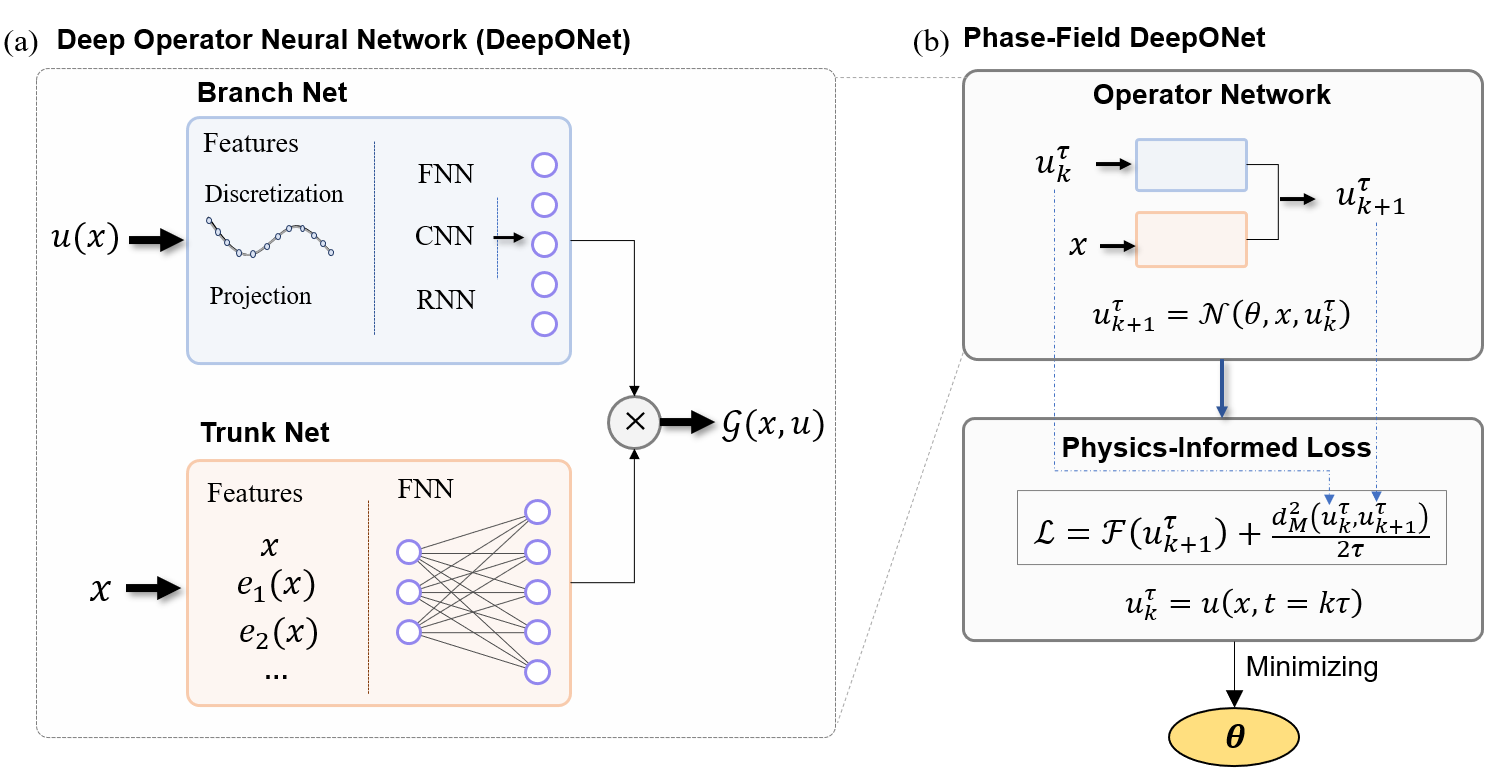}
    \caption{Schematic illustration of (a) DeepONet and (b) Physics-informed DeepONet with energy-based loss function.}
    \label{fig:deep_onet}
\end{figure}

DeepONet was first proposed by Lu et al. \cite{lu2019}. 
It is a high-level network structure with two sub-networks, namely the ``branch" network and the ``trunk" network, as shown in Figure \ref{fig:deep_onet}a. 
The trunk network takes the coordinates $x$ as the input, while the branch network takes the function $u$ as the input. 
The final output is given by multiplying the outputs of both networks,

\begin{equation}
    \label{eq:DeepONet_out}
    \mathcal{G}(x, u) = \sum_{k=1}^p {b_k(u) t_k(x)} + b_0,
\end{equation}

\noindent where $b_k $ and $ t_k (k=1, 2, ..., p)$ are the outputs of the branch network and trunk network, respectively; $p$ is the number of outputs of both sub-networks.

There are several important features we should note about DeepONet.
First, since the trunk network takes coordinates as input, the output is continuous, which means we can get predictions at any location.
More importantly, the gradients of outputs with respect to inputs can be easily estimated by automatic differentiation. This is crucial to constructing the physics-informed loss function. 
Second, it is highly flexible. The essence of this structure is to separate the vector input and function input into two sub-networks.
It can be tailored to fit different applications. 
On the one hand, feature expansion can be performed on the inputs. 
For the branch network, instead of a vector of discretized points of a function, we can extract any other features from the function to be the inputs, e.g, magnitude and phase in the frequency domain. 
On the other hand, the sub-networks can be any type of neural network such as the FNN, CNN, recurrent neural network (RNN), etc. 

DeepONet enables us to map one function to another. 
In real applications, the input function can be the initial or boundary condition and the output can be the solution at a random time. 
Here, we aim to predict the distribution of a field variable at any given time step. 
To achieve this, we propose to take the current distribution of a field variable as the input and the distribution at the next time step as the output (i.e. a mapping $u_k^\tau \rightarrow u_{k+1}^\tau$), which can be expressed by,

\begin{equation}
    \label{eq:Time_Stepper_Mapping}
    u_{k+1}^\tau = \sum_{k=1}^p {b_k(u_k^\tau) t_k(x)} + b_0,
\end{equation}

\noindent where $u_k^\tau = u(x, t=k\tau)$ represents the distribution of $u$ at the $k^\text{th}$ time step with a time interval of $\tau$.  
The proposed network structure works like an explicit time-stepper. Given the input at the $k^\text{th}$ time step $u_k^\tau$, we can get the output at the $(k+1)^\text{th}$ time step $u_{k+1}^\tau$, which can then be treated as the input to get $u_{k+2}^\tau$. 
Following this iterative process, a sequence of distributions $[u_k^\tau, u_{k+1}^\tau, u_{k+2}^\tau, ...]$ at all following time steps can be obtained. 

\subsection{Phase-Field DeepONet}

Machine learning algorithms often require a large dataset for training and the loss function is commonly constructed based on the error (e.g., mean square error, mean absolute error) between the data and predictions. 
Physics-informed machine learning, however, can work with only a small dataset or even without data. 
This is achieved by introducing physical laws (e.g. PDEs) into the loss function. 
Physics-informed machine learning can deal with two typical types of problems, namely the inverse problem and the forward problem. 
For the former one, the physics is partially known with a small dataset available, and the aim is to learn the unknown physics; 
for the latter one, we know all the physics without any data and the goal is to solve the governing equations. 
We will focus on the forward problem in this study. That is to solve the systems governed by gradient flows.

Herein, we propose a general framework with physics-informed DeepONet for gradient flows of free energy potential, as illustrated in Figure \ref{fig:deep_onet}b, which will be referred to as ``Phase-Field DeepONet". 
We include the DeepONet structure as mentioned above
and construct a physics-informed loss function according to the minimizing movement scheme (Eq.~\ref{eq: Min. Move. Scheme}).
The loss function can be written as,

\begin{equation}
    \label{eq:Loss_Func_General}
    \mathcal{L} = \mathcal{F}(u_{k+1}^\tau) + \frac{d^2(u_{k+1}^\tau, u_{k}^\tau)}{2 \tau}.
\end{equation}

The training process will minimize this loss function, equivalent to the minimizing movement scheme, therefore approximating the ground truth. 
To train this network, space coordinates $x$ and the distribution $u_k^\tau$ can be randomly sampled as the input of training data. 
There is no need for the output $u_{k+1}^\tau$ for the training. 
The trained DeepONets can work as efficient surrogates (explicit time-steppers) that are able to predict the time sequence of field distributions at all following time steps, given the current field distribution. The detailed training process will be explained in the following numerical examples.

It is worth noting that a PINN can be regarded as a special example of a physics-informed DeepONet, wherein the branch net is omitted or the input for the branch network is held constant as the initial condition. For example, if we only keep the trunk net in Figure \ref{fig:deep_onet}a and replace $u_k^\tau$ in Eq.~\ref{eq:Loss_Func_General} with $u_0^\tau$, the resulting framework is a PINN to solve for the distribution of $u$ in the next step.

\section{Numerical examples}\label{sec:numerical_examples}

\subsection{Relaxation kinetics}

We start with an example of the relaxation kinetics in 1D space governed by gradient flows. The governing free energy is

\begin{equation}
    \label{eq:Free_Energy_Relax_Kinetics}
    \mathcal{F} (u) = \int_{\Omega} \frac{1}{2}k \cdot u^2(x,t) \dif x.
\end{equation}

We consider a 1D domain within $[-1, 1]$. With the $L^2$ inner product, the corresponding PDE and boundary conditions can be derived with Eq.~\ref{eq:gradient_flow}, Eq.~\ref{eq:functional_derivative}, and Eq.~\ref{eq:L2_Allen-Cahn} ($M=1$),

\begin{equation}
    \label{eq:Relax_Kinetics_PDE_BC}
    \left\{
    \begin{array}{ll}
         \dpd{u}{t} = - k u,  \\
         u(x=\pm 1) = 0.
    \end{array}
    \right.
\end{equation}

Given the initial condition $ u_0 = u(x, t=0)$, an analytical solution can be found as $ u = - e^{k t} u_0 $. 
In this example, we will compare the two different frameworks, namely, the traditional PINN and the proposed Phase-Field DeepONet, 
to illustrate the feasibility of incorporating the minimizing movement scheme into a machine learning framework and also to discuss the advantages of the proposed framework.

\begin{figure}[h]
    \centering
    \includegraphics[width = \textwidth]{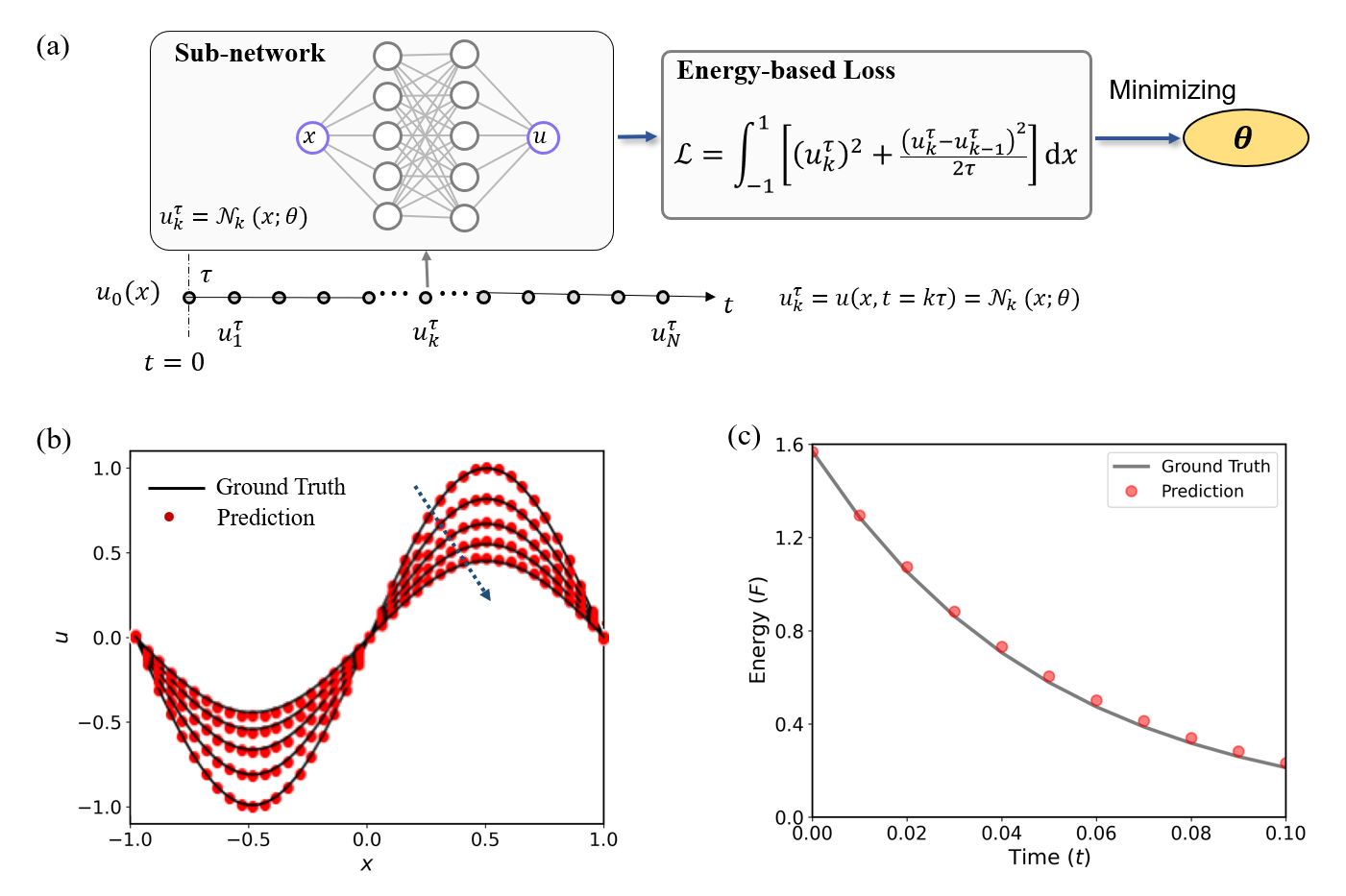}
    \caption{(a) The multi-network structure for minimizing movement scheme; (b) Comparison between the ground truth and neural network predictions at different time steps; (c) Predicted free energy change with time.}
    \label{fig:pinn_rela_kinetics}
\end{figure}

For the PINN framework, the initial condition needs to be given, which is set as $u_0= \text{sin}(\pi x)$. The constant $k$ is set to 10.
Following the minimizing movement scheme, a straightforward approach is to discretize the time domain into many time steps with a fixed time interval $\tau$. 
As shown in Figure \ref{fig:pinn_rela_kinetics}a, for each time step $k$, we construct one corresponding neural network $\mathcal{N}_k(x;\theta)$, where $x$ denotes the input spatial coordinate and $\theta$ represents the weights and biases to be optimized. 
The output of the sub-network is the current distribution of the field variable $u_k^\tau=u(x, t=k\tau)$. 
The sum of free energy and distance as in Eq.~\ref{eq: Min. Move. Scheme} is directly treated as the loss function,

\begin{equation}
    \label{eq:loss_PINN_relax}
    \begin{split}
     	\mathcal{L} &= \mathcal{F}(u_{k}^\tau) + \frac{d_{L^2}^2 \left( u_{k}^\tau, u_{k-1}^\tau \right) }{2 \tau} \\
                        &= \int_{-1}^1 \left[ (u_k^{\tau})^2  + \frac{(u_k^{\tau} - u_{k-1}^{\tau})^2}{2 \tau} \right] \dif x\\ 
                        &\approx \frac{1}{N_F}\sum_{i=1}^{N_F} (u_k^{\tau})^2 + 
                        \frac{1}{N_d}\sum_{i=1}^{N_d}  \frac{(u_k^{\tau} - u_{k-1}^{\tau})^2}{2 \tau},
    \end{split}
\end{equation}

\noindent where $u_k^{\tau,i} = u(x = x_i, t = k\tau)$ represents the field value at the sampled locations $x_i$; $N_F, N_d$ are the total number of samples to evaluate the numerical integration of the free energy term and distance term, respectively. $N_F$ and $N_d$ can be different; for simplicity, they are set to be equal in this study.
The training process follows the minimizing movement scheme: a) we first train the 1st sub-network $\mathcal{N}_1$ given the initial condition $u_0$; 
b) after training, we can get the output $u_1^{\tau}$. We then train the 2nd sub-network $\mathcal{N}_2$ with $u_1^{\tau}$ as the input; 
c) we repeat this process to train all the sub-networks sequentially.

All the sub-networks were constructed with two hidden layers with 20 nodes each. The ReLU activation function was adopted. 
We randomly sampled 1000 uniformly distributed data points within the space domain ($N_F = N_d = 1000$) as the training data. 
A learning rate of 0.001 and the Adam optimizer were used. 
Each sub-network was trained for 500 epochs in sequence and the training was repeated for 3 rounds. 
The training took around 25 minutes on a single GPU (NVIDIA T400 4GB) system.
The predictions of the sub-networks at different time steps are compared with the analytical results and a good match can be observed (Figure \ref{fig:pinn_rela_kinetics}b). 
We can also see the decreasing free energy of the system (Figure \ref{fig:pinn_rela_kinetics}c). 
These successfully validate the feasibility to implement the minimizing movement scheme into physics-informed machine learning. However, this PINN-based framework is  not very efficient.
First, all the sub-networks have to be trained in sequence and the number of sub-networks will increase largely if predictions at a large time interval are required. 
Second, the PINNs need to be retrained once the initial condition is changed, which limits their applications. 

\begin{figure}[h]
    \centering
    \includegraphics[width = \textwidth]{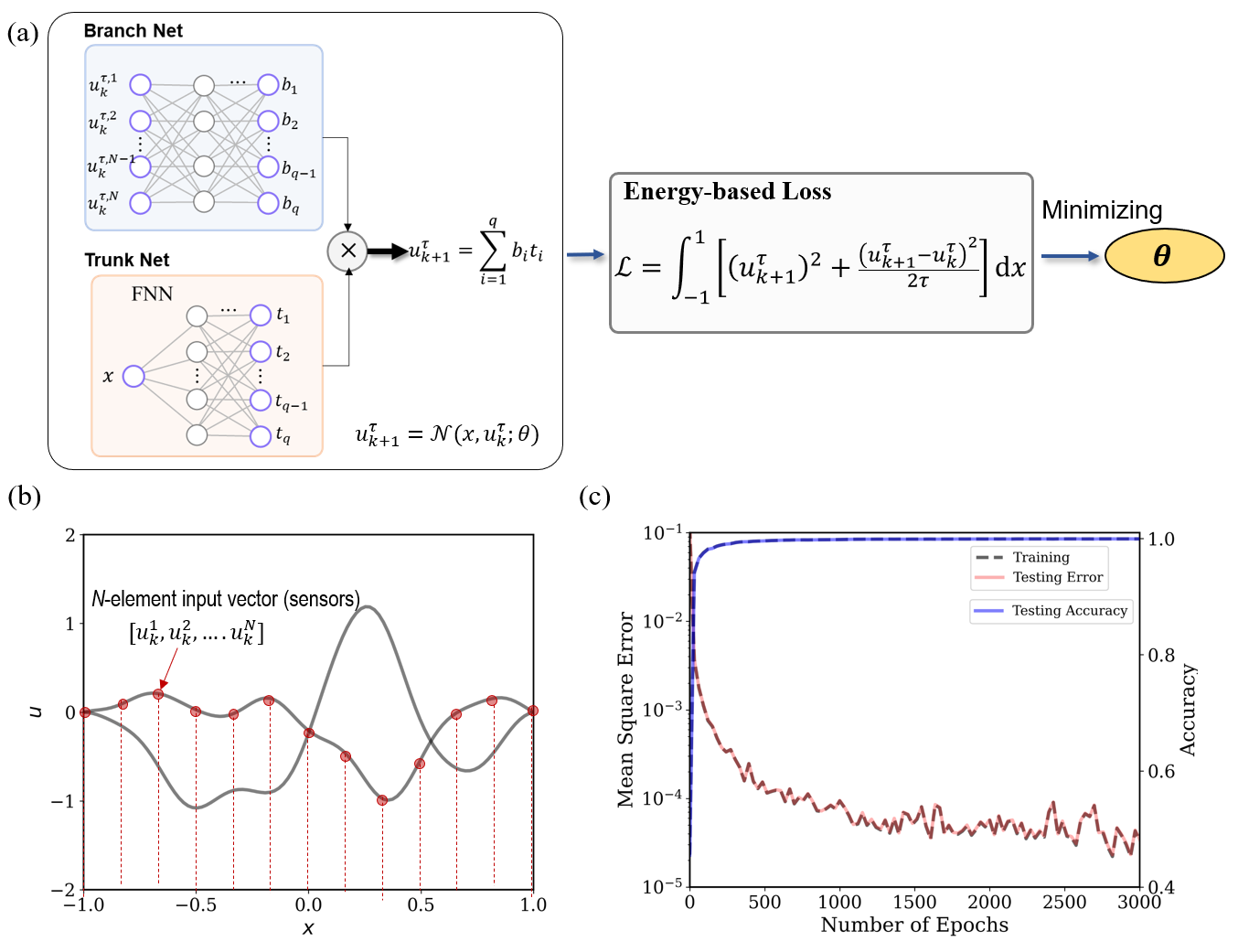}
    \caption{(a) Illustration of physics-informed DeepONet incorporating minimizing movement scheme for 1D relaxation kinetics; (b) Random distribution of $u$ generated by Gaussian Random Process and the discretized input vector; (c) Mean square error and r2-value (accuracy) in training and testing sets.}
    \label{fig:onet_rela_kinetics}
\end{figure}

To address these issues, as we proposed in the previous section, a DeepONet is constructed to take the distribution of the current time step as the input and predicts the distribution of the next time step, which works like an explicit time-stepper. 
As shown in Figure \ref{fig:onet_rela_kinetics}a, both the branch net and the trunk net are fully connected neural networks. 
Note that a continuous function cannot be directly fed into the branch net; discretization is therefore performed here. 
We discretized $u$ by sampling its values at equally spaced locations in the space domain; the resulting vector of values was then treated as the input (Figure \ref{fig:onet_rela_kinetics}b). 
The loss function is the same as in Eq.~\ref{eq:loss_PINN_relax}, except that $k$ and $k-1$ should be replaced by $k+1$ and $k$, respectively. 

\begin{table}[]
\label{tab:network_structure}
\caption{Paramters of the network structures for the three different cases}
\begin{tabular}{l|cc|cc|c}
\hline
\textbf{Case}            & \multicolumn{2}{c|}{\textbf{Trunk Net}}                                                                                                                                                                & \multicolumn{2}{c|}{\textbf{Branch Net}} & \textbf{\# of Sensors} \\ \hline
                         & Depth                                                                                                             & Width                                                                              & Depth               & Width              &                        \\
Relaxation Kinetics (1D) & 3                                                                                                                 & 100                                                                                & 2                   & 100                & 100                    \\
Cahn-Hilliard (1D)       & 3                                                                                                                 & 40                                                                                 & 2                   & 40                 & 40                     \\ \hline
Allen-Cahn (2D)          & \multicolumn{1}{l}{\begin{tabular}[c]{@{}l@{}}Filters:    32   \\ Kernels: (3,3)\\ Strides:  (1, 1)\end{tabular}} & \multicolumn{1}{l|}{\begin{tabular}[c]{@{}l@{}}6\\ (3,3)   \\ (3, 3)\end{tabular}} & 3                   & 120                & 120                   
\end{tabular}
\end{table}

\begin{figure}[h]
    \centering
    \includegraphics[width = \textwidth]{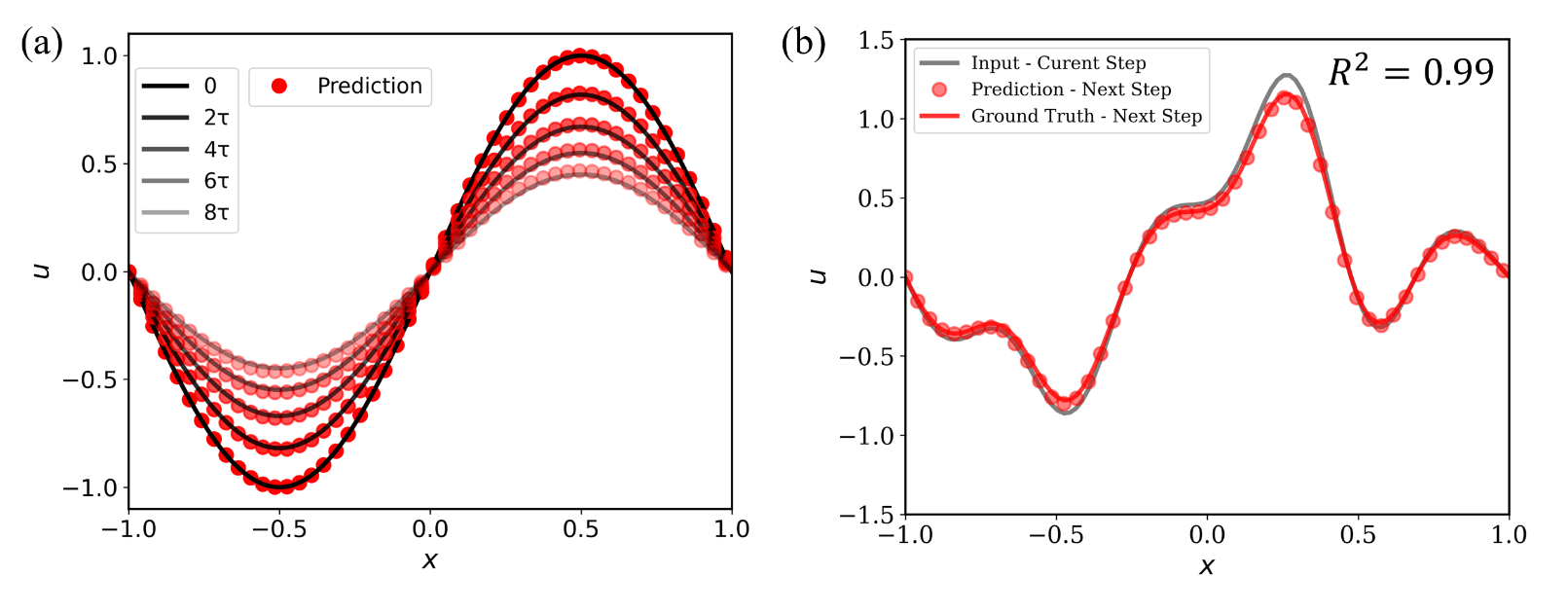}
    \caption{Predictions of trained DeepONet: (a) predictions at multiple time steps; (b) prediction for a random input.}
    \label{fig:onet_rela_kinetics_res}
\end{figure}

The input of the whole network is $[x, u_k^\tau]$. The physical constant $k$ is also set to 10; a given initial condition is unnecessary for DeepONet.
To simplify the training process, we evenly sampled 100 points ($N_F=N_d=100$) in the space domain [-1, 1]. 
We generated 10,000 different distributions of $u$ by  Gaussian random process, half of which was for training and the other half for testing. 
The size and structure of the DeepONet can be found in Table~\ref{tab:network_structure}. 
The network was trained for 3000 epochs with a learning rate of 0.001. The training time was around 10 minutes on the same single GPU system.
The training and testing error is shown in Figure \ref{fig:onet_rela_kinetics}c. 
The coefficient of determination (r2-value)  reached 0.99 in the testing set. This implies that the trained DeepONet can accurately predict the relaxation kinetics. 
Figure \ref{fig:onet_rela_kinetics_res}a demonstrates the predicted distributions of $u$ at different time steps, which agree well with the ground truth. 
Note that the sinusoidal input of $u$ is outside of the training dataset, which indicates a good generality of the trained DeepONet.
More importantly, given a random input of $u$ at any time step, the trained network can accurately predict the distribution of $u$ at the next step (Figure \ref{fig:onet_rela_kinetics_res}b).

\subsection{Allen-Cahn equation}

In this example, the more complex Allen-Cahn equation in 2D domain is explored. The total free energy governing this equation is,

\begin{equation} 
    \label{eq:allen-cahn_free_energy}
    \mathcal{F} = \int_\Omega \left [ \frac{1}{\varepsilon^2} f(u)  + 0.5(\nabla u)^2 \right] \dif A,
\end{equation}

\noindent where $f(u)$ is the bulk energy density and a common choice is $f(u) = \frac{(u^2 - 1)^2}{4} $; the second term represents the interracial energy. With the $L^2$ norm (or inner product),  the corresponding PDE and boundary conditions can then be derived with Eq.~\ref{eq:gradient_flow}, Eq.~\ref{eq:functional_derivative}, and Eq.~\ref{eq:L2_Allen-Cahn},

\begin{equation}
    \label{eq:allen_canh_2d}
     \left\{
    \begin{array}{ll}
         \dpd{u}{t} = \nabla^2u - \frac{1}{\varepsilon^2} \dod{f}{u},  \\
         \nabla u \cdot \mathbf{n} = 0.
    \end{array}
    \right.
\end{equation}

\begin{figure}
    \centering
    \includegraphics[width = \textwidth]{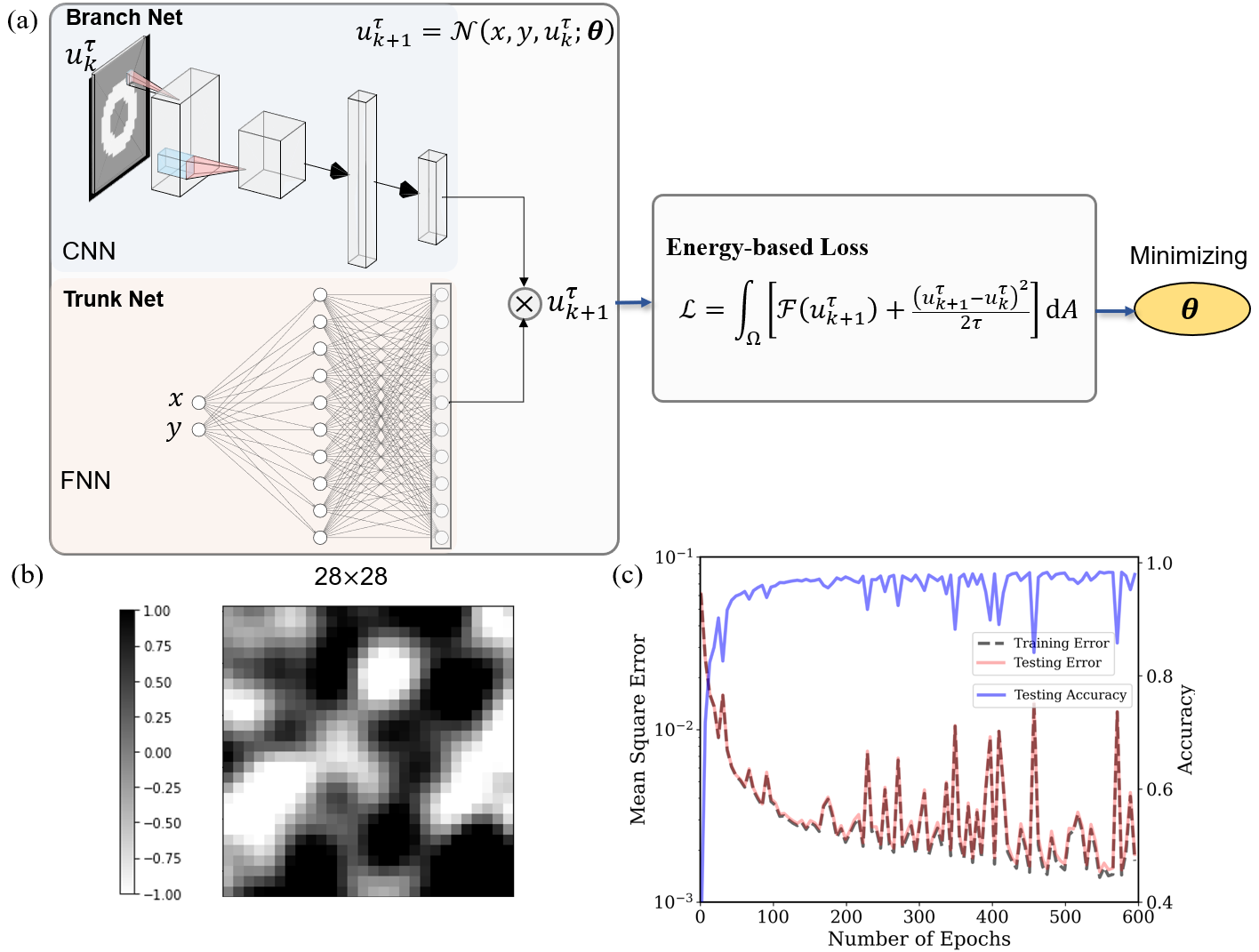}
    \caption{(a) Illustration of physics-informed DeepONet incorporating minimizing movement scheme for 2D Allen-Cahn equation; (b) Image input of the branch net ($28\times28$ grid of random distributed $u$ generated by Gaussian Random Process); (c) Mean square error and r2-value (accuracy) in training and testing sets.}
    \label{fig:onet_allen_cahn}
\end{figure}

A 2D domain $[-1,1]  \times [-1,1] $ is considered. The only physical constant, length scale $\epsilon$, is set to 0.25. 
We constructed a  DeepONet structure, as shown in Figure \ref{fig:onet_allen_cahn}a. 
The branch net takes the 2D distribution at the current time step as the input; a convolution neural network (CNN) is therefore adopted, which is then connected to a FNN. 
The trunk net is a FNN taking the spatial coordinates as the input. 
The final output is the distribution at the next time step.
The loss function for this case is 

\begin{equation}
    \label{eq:loss_AC}
    \begin{split}
     	\mathcal{L} &= \mathcal{F}(u_{k+1}^\tau) + \frac{d_{L^2}^2 \left( u_{k+1}^\tau, u_k^\tau \right) }{2 \tau} \\
                        &= \int_0^1\int_0^1 \left[ (\mathcal{F}(u_{k+1}^\tau)  + \frac{\left( u_{k+1}^{\tau} - u_k^{\tau} \right) ^2}{2 \tau} \right] \dif x \dif y\\ 
                        &\approx \frac{1}{N_{F}}\sum_{i=1}^{N_{F}} \mathcal{F}(u_{k+1}^{\tau,i}) + 
                        \frac{1}{N_{d}}\sum_{i=1}^{N_{d}}  \frac{( u_{k+1}^{\tau,i} - u_k^{\tau,i}) ^2}{2 \tau} ,
    \end{split}
\end{equation}

\noindent where $u_k^{\tau,i} = u(x = x_i, y = y_i, t = k\tau)$ denotes the field value at the sampled locations; 
 $N_F$ and $N_d$ are the total number of samples for evaluating the integral of the free energy and the distance term, respectively.
To train this network, random 2D distributions of $u$ are generated by Gaussian random process on a uniform $28\times28$ grid ($N_{F} = 256$) as the input of the branch net (Figure \ref{fig:onet_allen_cahn}b).
A total of 10,000 distributions (images) were generated, half of which were used for training and the remaining half for testing.
To simplify the training process, the corresponding spatial coordinates of the same uniform $28\times28$ grid ($N_{d} = N_{F} = 256$) were taken as the input of the trunk net. 
The time step $\tau = 0.005$.
The size and structure of the network can be found in Table~\ref{tab:network_structure}. The CNN in the branch net consists of two layers. 
The first layer has 32 filters with a kernel size of 3 and a stride of 1; the second layer has 6 filters with a kernel size of 3 and a stride of 3. 
The output of CNN is then flattened and fully connected to an output layer of 120 nodes. The trunk net consists of three hidden layers and each has 120 nodes.
The training process took around 150 minutes on the single GPU system due to the relatively large network and higher gradient terms in the loss function.

\begin{figure}
    \centering
    \includegraphics[width = \textwidth]{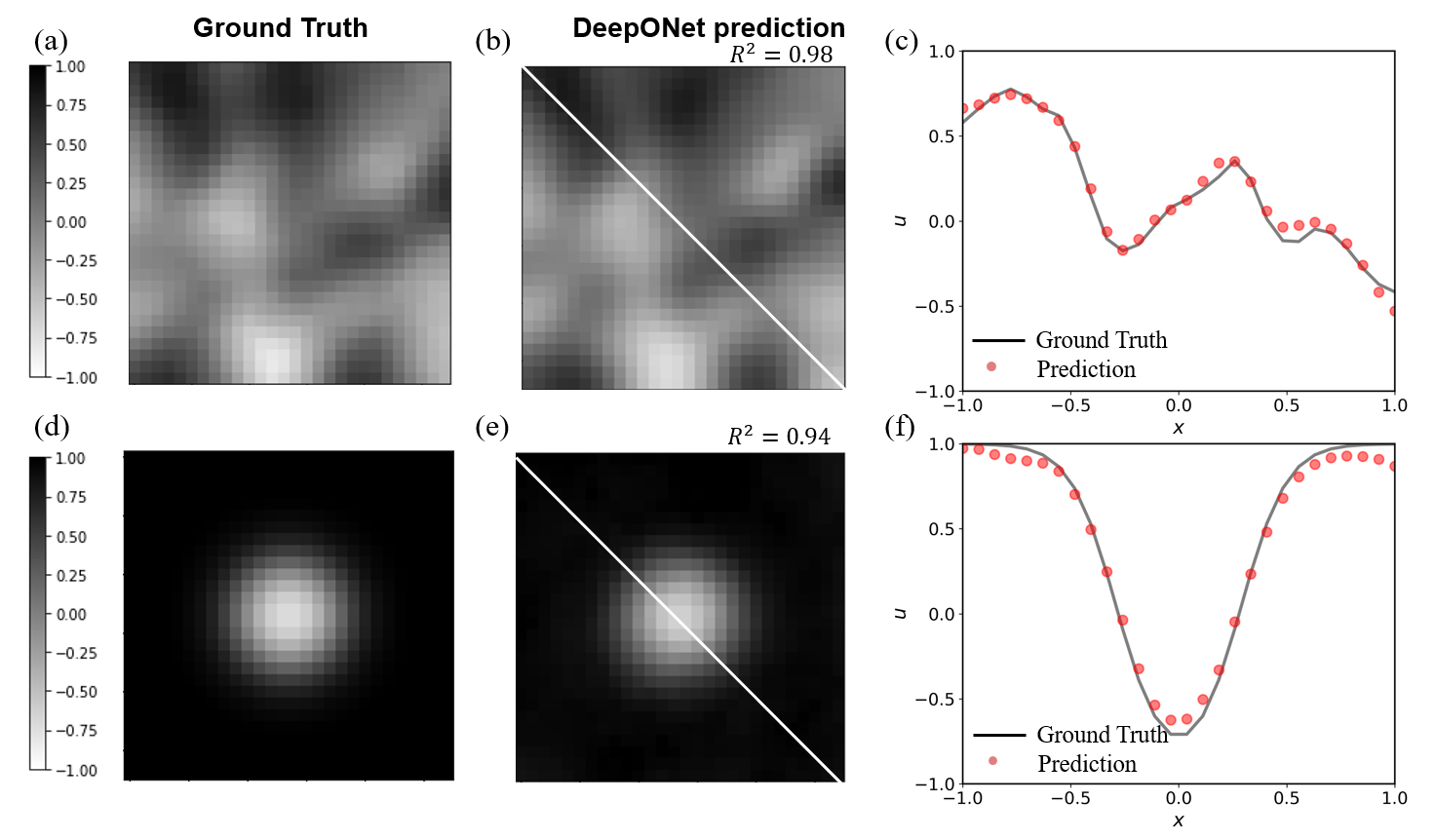}
    \caption{Comparison between the predictions of the trained DeepONet and the ground truth for two representative cases: (a), (b), (c) for a case in the testing set; (d), (e), (f) for another distinct case outside the dataset. Data points in (c) and (f) were extracted from the diagonals as indicated in (b) and (e). }
    \label{fig:onet_allen_cahn_res}
\end{figure}

Figure \ref{fig:onet_allen_cahn}c shows the training and testing error. An r2-value as high as 0.96 can be reached in the testing dataset. 
The predictions of the trained DeepONet are compared with the corresponding solution by the finite difference method, which is treated as the ground truth (Figure \ref{fig:onet_allen_cahn_res}). 
One representative sample in the testing set is compared in Figure \ref{fig:onet_allen_cahn_res}a, b, and c, where we can see the prediction of the neural network well matches the ground truth. 
We also compared a quite distinct case (Figure \ref{fig:onet_allen_cahn_res}d,e, and f), the distribution of which is not seen in the training set.
A good match can still be observed. This further validates the reliability of the proposed framework.

\subsection{Cahn-Hilliard equation}

An even more challenging case is the Cahn-Hilliard equation. Though the free energy is the same as in the previous case (Eq. \ref{eq:allen-cahn_free_energy}), a different inner product, the $H^{-1}$ inner product, is used to describe the dynamics, which gives a higher order of PDE with boundary conditions,

\begin{equation}
    \label{eq:canh_hilliard_1d}
     \left\{
    \begin{array}{ll}
         \dpd{u}{t} = \nabla^2\left( \frac{1}{\varepsilon^2} \dod{f}{u} - \nabla^2u \right) ,  \\
         \nabla \left( \frac{1}{\varepsilon^2} \dod{f}{u} - \nabla^2u \right) \cdot \mathbf{n} = 0, \\
         \nabla u \cdot \mathbf{n} = 0.
    \end{array}
    \right.
\end{equation}

We consider a 1D domain [0, 1]. The loss function is,

\begin{equation}
    \label{eq:loss_CH}
    \begin{split}
     	\mathcal{L} &= \mathcal{F}(u_{k+1}^\tau) + \frac{d_{L^2}^2 \left( u_{k+1}^\tau, u_k^\tau \right) }{2 \tau} \\
                        &= \int_0^1 \left[ (\mathcal{F}(u_{k+1}^\tau)  + \frac{( u_{k+1}^{\tau} - u_k^{\tau}) ( \phi_{k+1}^{\tau} - \phi_k^{\tau})}{2 \tau} \right] \dif x\\ 
                        &\approx \frac{1}{N_{F}}\sum_{i=1}^{N_{F}} \mathcal{F}(u_{k+1}^{\tau,i}) + 
                        \frac{1}{N_{d}}\sum_{i=1}^{N_{d}}  \frac{( u_{k+1}^{\tau,i} - u_k^{\tau,i}) ( \phi_{k+1}^{\tau, 1} - \phi_k^{\tau, 1})}{2 \tau} ,
    \end{split}
\end{equation}

\noindent where $\phi_k^\tau$ is the solution of Poisson's equation with the source term $u_k^\tau$. A linear mapping from discretized $u_k^\tau$ to $\phi_k^\tau$ can be obtained with the finite difference scheme,

\begin{equation}
    \label{eq:liear_mapping}
    \mathbf{\phi}_k^\tau = \mathbf{M}^{-1} \mathbf{u}_k^\tau,
\end{equation}

\noindent where $\mathbf{\phi}_k^\tau = [\phi_k^{\tau, 1}, \dots, \phi_k^{\tau, i}, \dots, \phi_k^{\tau, N_{d}}]$, $\mathbf{u}_k^\tau = [u_k^{\tau, 1}, \dots, u_k^{\tau, i}, \dots, u_k^{\tau, N_{d}}]$, and
\begin{equation}
    \label{eq:mapping_matrix}
    \mathbf{M} = 
    \frac{1}{\Delta x}
    \begin{bmatrix}
        -1 &  1     &        &         &        &        &   \\
         1 & -2     &  1     &         &        &        &   \\
           & \ddots & \ddots & \ddots  &        &        &   \\
           &        & 1      &  -2     & 1      &        &   \\
           &        &        & \ddots  & \ddots & \ddots &   \\
           &        &        &         & 1      & -2     & 1 \\
           &        &        &         &        & 1      &-1
    \end{bmatrix}
    , \Delta x = \frac{1}{N_{d} -1}.
\end{equation}

\begin{figure}
    \centering
    \includegraphics[width = \textwidth]{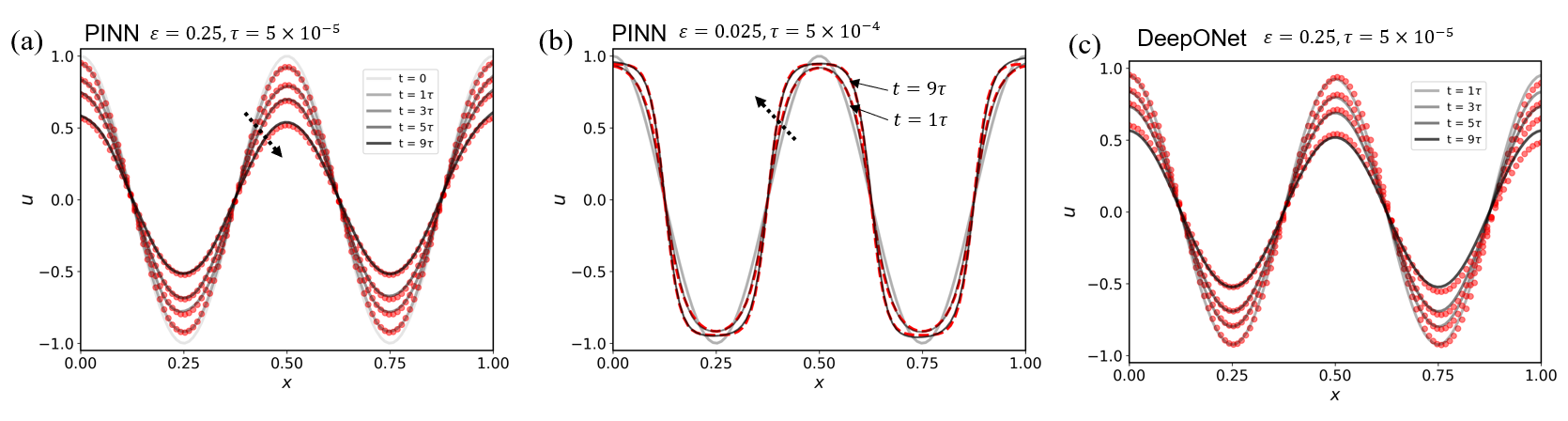}
    \caption{Predictions of PINN for cases without phase separation (a) and with phase separation (b). (c) Prediction of DeepONet for the case without phase separation.}
    \label{fig:cahn_hilliard_res}
\end{figure}

Considering the complexity of this problem, we started with a relatively simple PINN-based framework to further validate the feasibility of incorporating minimizing movement scheme into physics-informed machine learning.
An initial condition $u(x, t=0)=cos(4\pi x)$ and a boundary condition $u(x=0,t)=u(x=1,t)=0$ were given. 
We used the same structure as described in the first example. 
Two different cases with ($\epsilon = 0.025$) and without ($\epsilon = 0.25$) apparent phase separation were explored by varying the length scale $\epsilon$.
The corresponding time steps $\tau$ are $5\times10^{-4}$ and $5\times10^{-5}$ for the two cases.
The training process is the same as in the first example and the training results are shown in Figure~\ref{fig:cahn_hilliard_res}a, b. 
We can see a good agreement between the predictions of PINNs and the ground truth for both cases. 
This indicates that incorporating minimizing movement scheme into machine learning also works for this high $4^{\text{th}}$ order equation. 
 
 We then trained a DeepONet, the structure of which can be found in Table~\ref{tab:network_structure}. 
 Similarly, 10,000 randomly distributed $u$ were sampled for training (half) and testing (half). 
 Figure~\ref{fig:cahn_hilliard_res}c shows the results for the case without apparent phase separation. 
 The prediction of trained DeepONet can also well capture the ground truth at multiple time steps. 
 These further validate the feasibility of the proposed framework.
 It is important to note that accurately predicting the dynamic evolution of the phase separation remains a significant challenge with the current framework.
 This is mainly due to the relatively large time scale of phase separation and a small time step is still required to capture the fast phase separation process at the same time. 
 The slight difference with one time step makes it difficult to be captured by the current framework.
 We will further discuss this limitation in the next sections.

\section{Discussion}

So far, we successfully developed a general physics-informed deep operator neural network with an energy-based loss function and have validated it with three different numerical examples. Here, we would like to elaborate on a few aspects to deepen the theoretical base of this approach and broaden its applicability.

\subsection{Estimation of the time step $\tau$}
\begin{figure}
    \centering
    \includegraphics[width = \textwidth]{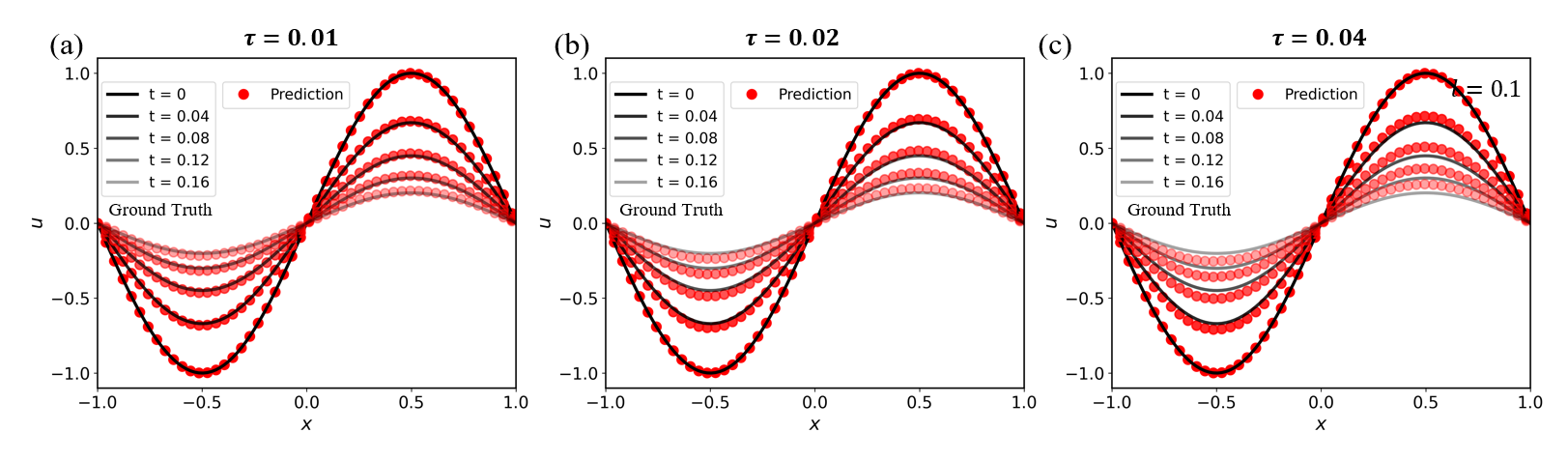}
    \caption{Predictions of DeepONets for cases with different time steps $\tau$:  (a) $\tau = 0.01$, (b) $\tau = 0.02$, (c) $\tau = 0.04$.}
    \label{fig:influence_time_step}
\end{figure}

The time step $\tau$ needs to be small enough to achieve an accurate solution. 
However, to the authors' best knowledge, there is no direct method to estimate the upper limit of $\tau$. 
In this study, we gradually decreased the time step to ensure the predictions converged to the ground truth. 
Figure \ref{fig:influence_time_step} shows the training results for three different values of $\tau$ in the kinetic relaxation example. 
As the value of $\tau$ increases from 0.01 (Figure \ref{fig:influence_time_step}a) to 0.02 (Figure \ref{fig:influence_time_step}b) and 0.04 (Figure \ref{fig:influence_time_step}c), a growing discrepancy between DeepONet predictions and ground truth can be observed. 

In practice, we have two suggestions for estimating the time step.
First, select a time step that enables the detection of a noticeable change in the field variable if experiment data is available.
Second, optimize this time step with a trial-and-error process based on PINN.
As we mentioned earlier, PINN is a simplified case of DeepONet with only the trunk net. Training PINN is much faster than training DeepONet and it is therefore suggested to determine $\tau$ with PINN. 
Since the estimation of the time step should be independent of the initial conditions,  a time step ensuring a converged solution for PINN should also work for DeepONet.

The requirement of a small time step is also a limitation of the proposed framework. 
When making predictions on a large time scale, the trained DeepONets need to be evaluated more times. 
This will decrease the efficiency and the error is expected to accumulate. 
To avoid this, we can train a DeepONet without including the minimizing movement scheme. The outputs are therefore required, either from experiments or simulations (supervised learning). 
This way, a large time step can be selected to further increase the computation efficiency.

The potential for implementing an adaptive time step within this framework is another compelling topic to investigate. 
As in the phase separation case of the Cahn-Hilliard equation (Figure~\ref{fig:cahn_hilliard_res}b), prior to the onset of phase separation, the field variable undergoes minimal changes and this transitional period occurs over a relatively extended temporal duration. 
In contrast, during phase separation, the process unfolds rapidly within a short time frame.
Therefore, the possible implementation of an adaptive time step, wherein the time step is larger during the pre-phase separation period and smaller during the phase separation period, has the potential to be advantageous in terms of computation efficiency.

\subsection{Influence of training samples}
\begin{figure}
    \centering
    \includegraphics[width = \textwidth]{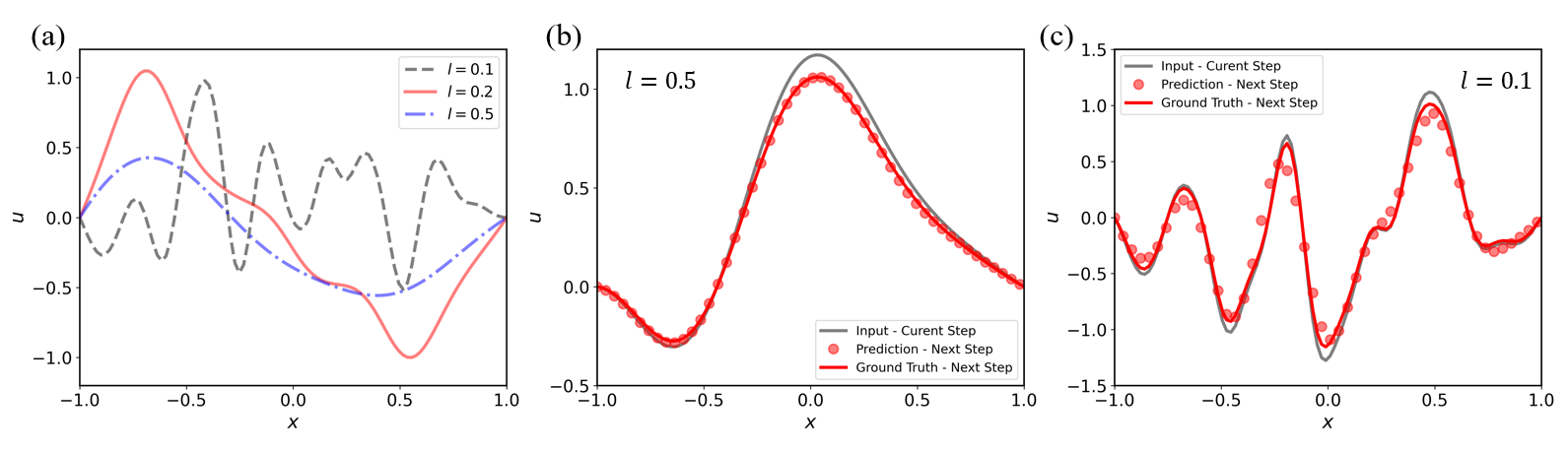}
    \caption{(a) Random distributions generated with different length scales $l = 0.1, 0.2, and 0.5$; predictions of DeepONets for randomly generated inputs of $u$ with length scales $l = 0.5$ (b) and  $l = 0.5$ (c).}
    \label{fig:influence_sampling}
\end{figure}

The sampling method is another possible factor influencing the accuracy. The input function $u(x)$ of the field variable is generated by a mean-zero Gaussian random process,
\begin{equation}
    \label{eq:GRP}
    \begin{split}
        u(x) & \sim \mathbf{G}(0,k_l(x_1, x_2)), \\
        k_l(x_1, x_2) & = exp(-||x_1-x_2||^2/2l^2),
    \end{split}
\end{equation}
where $k_l(x_1, x_2) $ is the covariance kernel with a length-scale parameter $l$. 
It controls the smoothness of the generated distributions of $u$. 
A larger $l$ results in a smoother $u$ (Figure \ref{fig:influence_sampling}a). 

In this study, the training and testing datasets are set to have the same ``smoothness" for simplicity. 
For example, $l = 0.2$ is used for both training and testing datasets in the first numerical example.
However, this may impact the accuracy when the input distribution of $u$ has a different degree of smoothness.
Figure \ref{fig:influence_sampling}b and c show the predictions of the same DeepONet with a smoother ($l=0.5$) and a sharper ($l=0.1$) input of randomly disturbed $u$, respectively.
A strong agreement can still be seen for the smoother input, however, discrepancies can be observed in the sharp corners for the sharper input.
One potential solution for further enhancing the performance is to include training data with varying degrees of smoothness.
Note that the overall trend is captured reasonably well even with the sharper input. This demonstrates the generality of this framework when applied to random inputs.

\subsection{Other potential improvements}

Apart from optimizing the time step and sampling dataset, there are several other aspects that could be improved in the future for this framework.

a) Accounting for a wider range of physical constants. This will be beneficial to practical applications. 
Currently, a fixed set of physical constants is used for simplicity, but in reality, these values may vary across different systems. 
To address this issue, physical constants could be included as inputs by adding an extra branch net, allowing a single DeepONet to be trained for all possible physical constants.

b) Handling irregular-shaped domains. This would further enhance its applicability. DeepONet is capable of handling any fixed irregular domain by randomly sampling locations in a 2D or higher dimensional space and feeding them into the network as a vector. However, the challenge is how  a DeepONet trained for a specific domain can be applied to domains with other geometric shapes.

c) Extending the framework to higher-dimensional problems. 
It would be interesting to investigate whether DeepONets trained at lower dimensions can be effectively applied to higher dimensions. 
Phase-field simulations at higher dimensions are notoriously computationally expensive, and scaling up the simulation using DeepONets could potentially overcome this issue.

d) Inverse learning from experiment data. The current framework focuses on the forward problem, where all physics and constants are known and the DeepONet approximates the solution with unsupervised learning (no experiment data needed).
It will be even more meaningful if we could extend the framework to the inverse problem. 
In the inverse problem, only part of the physics and some experimental data are known, and the goal is to identify the unknown physical constants or laws. 
Modifying the loss function to include the experimental data is one way to realize this. 
By doing so, both physics and experimental data can be included in one framework, and the training process is able to approximate the ground truth as well as learn the unknown physical constants and laws. 
However, the training process can be time-consuming and is not applicable for fast or real-time identification. 
Developing a framework for fast identification based on DeepONets would require more effort.

\section{Conclusion}

We propose a physics-informed Phase-Field DeepONet framework for dynamical systems governed by gradient flows of free energy functionals. 
The minimizing movement scheme is incorporated into the framework to solve the system dynamics instead of directly solving the governing PDEs. 
Three different numerical examples validate the proposed framework, including the two major equations of the phase-field method, namely the Allen-Cahn and Cahn-Hilliard equations. Some major conclusions can be drawn from this work:

\begin{enumerate}
    \item Variational principles, such as gradient flows, hold great potential to be seamlessly integrated into a physics-informed machine learning framework, providing a novel approach for the fusion of data and physics with wider practical implications.
    \item The proposed Phase-Field DeepONet framework successfully solves both the Allen-Cahn and Cahn-Hilliard equations in the phase-field method, demonstrating its effectiveness in simulating pattern formation in chemical systems.
    \item The Phase-Field DeepONets trained in this study can serve as efficient explicit time-steppers, potentially enabling fast real-time predictions of dynamic systems.
\end{enumerate}

This work raises the possibility of deep operator learning of more general phase-field models, including those of chemical nonequilibrium thermodynamics~\cite{bazant2011,Bazant2017}, which involve nonlinear dependencies of fluxes or reaction rates on diffusional chemical potentials. The minimizing movement scheme would need to be extended to go beyond the gradient flows approximation to account for nonlinear dynamics. In this way,  the Phase-Field DeepONet framework could enable data-driven learning and fast simulations of pattern formation from rich image datasets, going beyond PDE-constrained optimization~\cite{zhao2022learning}.

\section*{Acknowledgment}
W.L. and J.Z. gratefully acknowledge the support of the present work through the NASA 19-TTT-0103 project (Award No. 80NSSC21M0114). They are also supported by the Northeastern University and College of Engineering startup funds. M.Z.B and W.L. are grateful for the support of Toyota Research Institute through the D3BATT Center on Data-Driven-Design of Rechargeable Batteries.

\begin{appendices}[title]
\section{Basics of variational calculus}
\subsection{Euler-Lagrangian equation}\label{sec:appendix_EL}

 Given a smooth manifold $X$ and a smooth real-value function $f=f(x, u(x), u'(x)) $, the functional $\mathcal{F}$ defined as
 \begin{equation}
     \label{eq:Euler_Lagrange_functional}
     \mathcal{F} = \int_{x_a}^{x_b} {f(x, u(x), u'(x))} \dif x
 \end{equation}
 has a stationary value (maximum, minimum, or saddle point) if the Euler-Lagrangian equation is satisfied,
 
 \begin{equation}
     \label{eq:Euler_Lagrange_equation}
     \dpd{f}{u_i} - \dod{}{x} \dpd{f}{(u_i)} = 0, \quad i=1, 2, ..., n.
 \end{equation}
 
\subsection{Derivation of functional derivative}\label{sec:appendix_func_grad}

Consider the functional derivative of a specific type of energy functional that only depends on the field variable and its first-order derivatives, i.e., $ \mathcal{F} = \int_\Omega {F(x, u(x), \nabla u(x)} \dif x $. We have

\begin{equation}
    \label{eq: energy functional direvative}
    \begin{split}
        \int_\Omega {\frac{\delta \mathcal{F}}{\delta u} v} \dif x &= \left[ \dod{}{t} \int_\Omega {F(x, u + v t, \nabla u + \nabla v t)} \dif x \right]_{t=0} \\
        & = \int_\Omega \left( \dpd{F}{u} v + \dpd{F}{\nabla u} \cdot \nabla v \right) \dif x \\
        & = \int_\Omega \left( \dpd{F}{u} v - (\nabla \cdot \dpd{F}{\nabla u})  v \right) \dif x + \int_{\partial \Omega} {\left( \dpd{F}{\nabla u} \cdot \mathbf{n} \right) v}\dif x \\
        & = \int_\Omega \left( \dpd{F}{u}  - \nabla \cdot \dpd{F}{\nabla u} \right) v \dif x.
    \end{split} 
\end{equation}

The fourth line is valid when $v=0$ or $\dpd{F}{\nabla u} \cdot \mathbf{n} = 0$ on the boundary. Since $v$ is an arbitrary function, we get the functional derivative
\begin{equation}
    \label{eq:full_functional_derivative}
    \frac{\delta \mathcal{F}}{\delta u} = \dpd{F}{u}  - \nabla \cdot \dpd{F}{\nabla u}.
\end{equation}

\bibliographystyle{unsrt}  
\bibliography{references, Zhao}    

\end{appendices}
\end{document}